\documentclass[prd,preprint,showkeys,nofootinbib,a4paper]{revtex4-1}
\usepackage{amsmath}
\usepackage{amssymb}
\usepackage{mathrsfs}
\usepackage{graphicx}
\usepackage{bbm}
\usepackage[]{hyperref}

\def\eq#1 { \begin{equation} #1 \end{equation} }
\def\eqn#1{ \begin{eqnarray} #1 \end{eqnarray} }
\def\nn { \nonumber }

\def\Re{{\rm Re}\,}
\def\Im{{\rm Im}\,}
\def\Reals{\mathbb{R}}
\def\Nat{\mathbb{N}_0}

\def\half{\frac{1}{2}}

\def\calA{\mathcal{A}}
\def\a{\alpha}
\def\l{\lambda}
\def\s{\sigma}
\def\r{\rho}
\def\D{\Delta}
\def\Dr{\Delta^{\rm reg}}
\def\psir{\psi^{\rm reg}}
\def\eps{\epsilon}
\def\cM{\mathcal{M}}
\def\cP{\mathcal{P}}
\def\cV{\mathcal{V}}

\def\ket#1{\vert #1 \rangle}

\def\brak#1#2#3{\langle #1 \vert #2 \vert #3 \rangle}
\def \fvac {\ket{0}_{\rm free}}
\def\C#1{\left\langle #1 \right\rangle}

\def\G#1{\Gamma\left(#1\right)}
\def\GG#1{ \Gamma\left[ #1 \right] }

\def \GGG#1#2{\,\Gamma\left[ \begin{array}{l}
      #1 \\
      #2
    \end{array} \right]}
\def\2F1#1#2#3#4{\,\phantom{}_2F_1\left[#1\,,\,#2\,;\,#3\,;\,#4\right] }

\begin{document}
\title{The IR stability of de Sitter QFT: results at all orders}
\author{Donald Marolf}
\email{marolf@physics.ucsb.edu}
\author{Ian A. Morrison}
\email{ian\_morrison@physics.ucsb.edu}
\affiliation{University of California at Santa Barbara,
Santa Barbara, CA 93106, USA}

\begin{abstract} 
  We show that the Hartle-Hawking vacuum for theories
  of interacting massive scalars in de Sitter space is both perturbatively
  well-defined and stable in the IR.
  Correlation functions in this state may be
  computed on the Euclidean section and Wick-rotated to Lorentz-signature.
  The results are manifestly
  de Sitter-invariant and contain only the familiar UV singularities.
  More importantly, the connected parts of all Lorentz-signature
  correlators decay at large separations of their arguments.
  Our results apply to all cases in which the free Euclidean vacuum
  is well defined, including scalars with masses belonging to both the
  complementary and principal series of $SO(D,1)$.
  This suggests that interacting QFTs in de Sitter -- including
  higher spin fields -- are perturbatively IR-stable at least when
  i) the Euclidean vacuum of the zero-coupling theory exists and
  ii) corresponding Lorentz-signature zero-coupling correlators decay
  at large separations. This work has significant overlap with a paper 
  by Stefan Hollands, which is being released simultaneously.
\end{abstract}

\maketitle

\section{Introduction}
\label{sec:intro}

While free quantum fields in de Sitter space ($dS_D$) have been well understood for some time (see \cite{Allen:1985ux} for scalar fields), interacting de Sitter quantum field theory continues to be a topic of much discussion.  In particular, there has been significant interest in the possibility of large infrared (IR) effects in interacting de Sitter quantum field theories
\cite{Starobinsky:1979ty,Mottola:1984ar,Mottola:1991dc,Antoniadis:2006wq,Mottola:2010gp,Hu:1985uy,Hu:1986cv,Tsamis:1992sx,Tsamis:1994ca,Tsamis:1996qm,Tsamis:1996qq,Tsamis:2005hd,Higuchi:2000ge,Higuchi:2001uv,Higuchi:2002sc,
Polyakov:2007mm,PerezNadal:2008ju,Faizal:2008ns,Akhmedov:2008pu,Higuchi:2008tn,Higuchi:2009zza,Higuchi:2009ew,Akhmedov:2009ta,Polyakov:2009nq,Burgess:2010dd,Giddings:2010nc}, both with and without dynamical gravity.

In \cite{Marolf:2010zp} we began to address the specific class of such concerns associated with infra-red (IR) divergences of the naive Lorentz-signature de Sitter Feynman diagrams, or more generally those concerns that can be addressed in the context of minimally-coupled scalar fields with mass $M^2 > 0$.  There we computed one-loop corrections to propagators on Euclidean de Sitter (which is just the $D$-sphere $S^D$) and analytically continued the results to Lorentz-signature.
This procedure defines the so-called Hartle-Hawking vacuum of the Lorentzian theory \cite{Hartle:1976tp}, which on general grounds should be a good quantum state (see section \ref{sec:disc}).  In particular, the analytically continued correlators are expectation values of products of operators in a single state as opposed to matrix elements between an ``in-vacuum'' and a potentially different ``out-vacuum.'' We do not attempt to define any notion of S-matrix.

Because $S^D$ is compact, it is a priori clear that Euclidean correlators do not suffer infra-red divergences.  We showed in \cite{Marolf:2010zp} that, to one-loop order, the analytically continued Lorentz-signature correlators were also finite and decayed at a rate determined by the lightest relevant mass\footnote{In addition, the one-loop calculations reported in \cite{PerezNadal:2009hr} establish that correlators of free-field stress tensors decay at large separations.}.  The purpose of the current paper is to extend these results to arbitrary $N$-point functions and to all orders in perturbation theory, again showing that connected correlators decay rapidly as the separation between points becomes large.  As in \cite{Marolf:2010zp}, our results will apply to all masses for which the free Euclidean de Sitter vacuum is well-defined, i.e. for all $M^2 > 0$,   including values in both the complimentary series and the principal series of $SO(D,1)$.

The decay of connected correlators demonstrates that the Hartle-Hawking state is perturbatively stable, and that the Hartle-Hawking vacuum is an attractor state for local operators in the sense defined in \cite{Marolf:2010zp}.  To illustrate the main point, let us consider a state constructed from the Hartle-Hawking vacuum $|0 \rangle_{HH}$ with appropriately smeared operators:
\eq{ \label{eq:Psi}
  \ket{\Psi} := \int_{Y_1} \dots \int_{Y_n} f(Y_1,\dots,Y_n)
  \phi_{\s}(Y_1)\cdots \phi_{\s}(Y_n) |0 \rangle_{HH}.
}
Here the $Y_i$ are points in $dS_D$, $\int_Y\dots$ denotes an
integral over de Sitter, and $f(Y_1,\dots,Y_n)$ is a smearing function
which we assume to be supported in a compact domain $\mathcal{D}$.
Now examine the correlation function
$\brak{\Psi}{\phi_\s(X_1)\cdots\phi_\s(X_N)}{\Psi}$
with all $X_i$ at large separations from $\mathcal{D}$. In this
configuration the correlator is simply a smeared correlation
function between $2n$ operators located within $\mathcal{D}$ and $N$
operators with large (say, roughly equal) separations $|Z|$ from $\mathcal{D}$ evaluated in the Hartle-Hawking vacuum.  Since the associated connected correlators decay rapidly at large separations, this function approximately factorizes into a product of two correlators: one for the points in $\mathcal{D}$ and one for the other points.  The former factor is just the norm of $|\Psi \rangle$, so we have
$\brak{\Psi}{\phi_\s(X_1)\cdots\phi_\s(X_N)}{\Psi}
\to \langle \Psi | \Psi \rangle \cdot
{}_{HH}\brak{0}{\phi_\s(X_1)\cdots \phi_\s(X_N)}{0}{}_{HH}$.
This means that, as probed by local operators, the excited state $\ket{\Psi}$
becomes indistinguishable from the Hartle-Hawking vacuum.

We begin by briefly reviewing free de Sitter quantum field theory in section \ref{sec:free}.  We then address simple tree diagrams in section \ref{sec:Euclidean}, which also serves to introduce some useful Mellin-Barnes techniques and our choice of (Pauli-Villars) regularization scheme.  We address general diagrams in section \ref{sec:loops}, where  we establish the desired results for finite Pauli-Villars regulator masses (so that all diagrams are finite).  Since the infra-red asymptotics are independent of the regulator masses, it is straightforward to take the limit where such regulators are removed\footnote{\label{foot} After subtracting regulator-dependent local counter-terms in order to obtain a finite result.  We consider theories can be renormalized in this way. One would expect this procedure to be equivalent (up to finite local counter-terms) to the renormalization prescription given in \cite{Hollands:2001fb}, and thus to define a fully covariant renormalized quantum field theory in the sense of \cite{Hollands:2008vx} whenever the flat-space limit is power-counting renormalizable. However, we have not analyzed this question in detail and save any investigation for future work.}. Some technical material is relegated to the appendices.  We close with some discussion in section \ref{sec:disc}.

{\bf Remark:}  While paper was being prepared, we received a draft of \cite{Hollands:2010pr} which reports similar results.

\section{Free de Sitter QFT}
\label{sec:free}

This brief section serves as a review of scalar quantum field theory
in de Sitter and allows us to establish our notation.
We consider $D$-dimensional de Sitter space $dS_D$ with radius $\ell$,
which may be defined as the single-sheet hyperboloid in a $D+1$-dimensional
Minkowski space $M_{D+1}$. Points on de Sitter satisfy \cite{Birrell:1982ix}
\eq{
  \eta_{AB} X^A X^B = \ell^2 ,
}
where $X^A$ is a vector in the embedding space and
$\eta_{AB} = {\rm diag}(-1,1,\dots,1)$ is the usual Minkowski metric. Henceforth we will drop the index notation
and denote the inner product of two embedding space vectors $X_1$
and $X_2$ simply by $X_1 \cdot X_2$. For two points on de Sitter located at
$X_1$ and $X_2$ the inner product $X_1 \cdot X_2/\ell^2$ provides a convenient
measure of distance which we loosely call the \emph{embedding distance} between
$X_1$ and $X_2$ \cite{Allen:1985ux}. The embedding distance is related
to the length of the chord between $X_1$ and $X_2$ in the embedding space (with the length being proportional to $1-X_1 \cdot X_2$)
and is clearly invariant under the full de Sitter isometry group $SO(D,1)$.
The embedding distance satisfies:
\begin{itemize}
    \item $X_1 \cdot X_2/\ell^2 \in [-1,1)$ for spacelike separation,
    \item $X_1 \cdot X_2/\ell^2 =1$ for null separation, and
    \item $|X_1 \cdot X_2/\ell^2| > 1$ for timelike separation.
\end{itemize}
The antipodal point of $X_1$ is simply $-X_1$; clearly the embedding
distance between antipodal points is $-1$. See Figure~1.

\begin{figure}
  \label{fig:Z}
  \includegraphics[width=5.5in]{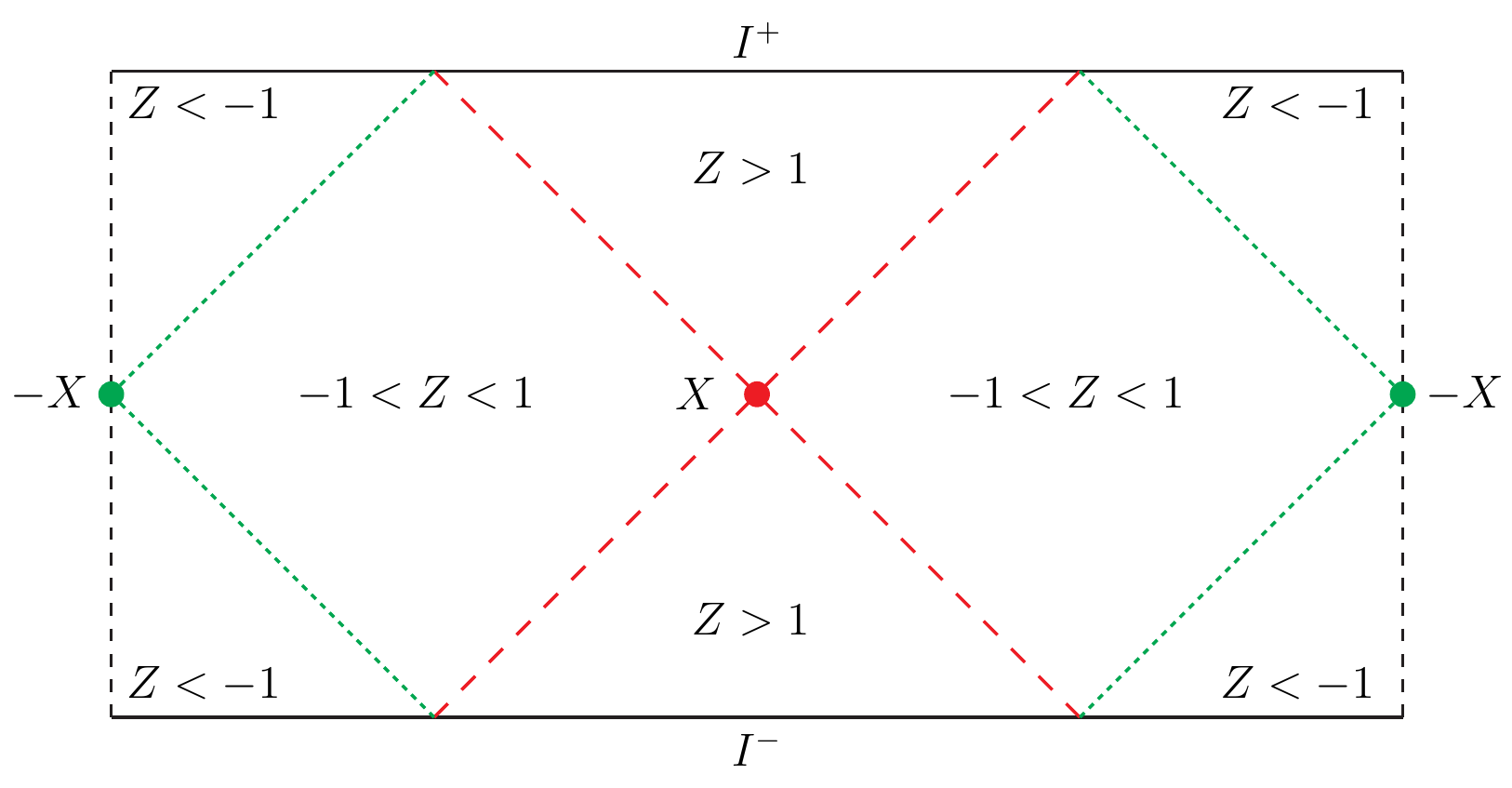}
  \caption{
    The conformal diagram of global de Sitter. The dashed
    ends are identified. Shown are the points $X$ and the corresponding
    antipodal point $-X$. Values of the embedding distance
    $Z := X Y/\ell^2$ in different regions of de Sitter are labeled;
    in addition, the dashed red lines denote the lightcone with $Z=1$
    and the dotted green lines denote the lightcone with $Z=-1$
    \cite{Allen:1985ux}.
  }
\end{figure}

In this work we restrict attention to massive scalar fields $\phi_\s(X)$.
It is convenient to keep track of the spacetime dimension with the
parameter $\a = (D-1)/2$; the mass parameter $\s$ is then defined by
the equation
\eq{ \label{eq:sigma}
  -\s(\s+2\a) = M^2\ell^2 ,
}
where $M^2$ is the bare mass-squared of the field if we assume minimal coupling to the metric.
There is a redundancy in this definition as (\ref{eq:sigma})
is invariant under $\s \to -(\s+2\a)$; for clarity we choose to
define $\s$ as the positive root
\eq{ \label{eq:sigma+}
  \s := -\a + \left(\a^2 - M^2\ell^2 \right)^{1/2} ,
}
but all expressions involving $\s$ must necessarily be invariant
under $\s \to -(\s+2\a)$. Free scalar fields form irreducible
representations of the de Sitter group $SO_0(D,1)$ and fall into three
series \cite{Vilenkin:1991a}:
\begin{enumerate}
  \item
    complementary series: $ -\a < \s < 0$ ,
  \item
    principal series: $\s = -\a + i \rho$, $\rho \in \Reals,\;\rho \ge 0$ ,
  \item
    discrete series: $\s = 0,1,2,\dots$ .
\end{enumerate}
We plot $\s$ and $-(\s+2\a)$ as a function of $M^2 > 0$ in 
figure~2.
~Relatively light massive fields belong to the
complimentary series while heavier fields belong to the principal series.  It is useful to note that $\sigma_{cc} = \frac{1}{2} - \alpha$ corresponds to an otherwise massless conformally coupled free field. This value lies in the complimentary series so long as $D > 2$.
Discrete series fields correspond to massless and tachyonic
scalars and we will not consider them here.

\begin{figure}
 \label{fig:sigma}
 \includegraphics[]{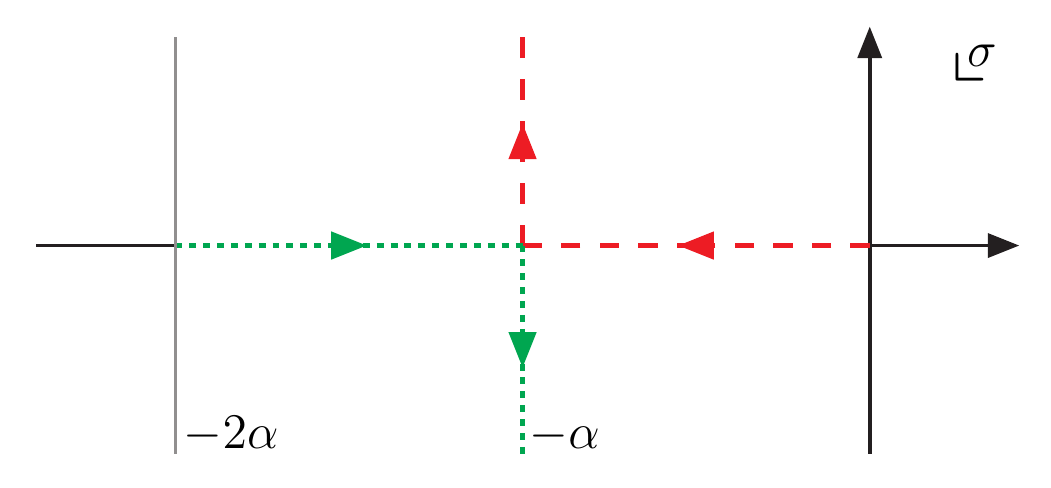}
  \caption{
    On-shell values of $\s$ and $-(\s+2\a)$ in the complex plane for
    massive scalar fields. The red dashed line denotes the path of $\s$
    for increasing $M^2$ starting from at $\s=0$ for $M^2 = 0$.
    The green doted line shows the path of $-(\s+2\a)$ for increasing $M^2$
    starting from $-(\s+2\a) = -2\a$ for $M^2 = 0$.
    Relatively light fields with $0 < M^2\ell^2 < \a^2$
    correspond to values of $\s$ and $-(\s+2\a)$ on the negative
    real axis and belong to the complementary series. Heavier
    fields with $M^2\ell^2 \ge \a^2$ correspond to complex values
    of $\s$ and $-(\s+2\a)$ on the line defined by
    $\Re \s = \Re(-\s-2\a) = -\a$.
  }
\end{figure}

Free massive scalar fields admit a unique de Sitter-invariant
Hadamard vacuum $\fvac$, commonly referred to as the Euclidean vacuum
(it is also the Bunch-Davies vacuum) \cite{Allen:1985ux,Mottola:1984ar}. Since the theory is free, the vacuum
is completely characterized by its 2-point functions. Let us
define the function
\eq{ \label{eq:Delta}
  \D_\s( Z ) :=
  \frac{\ell^{2-D}}{(4\pi)^{\a+1/2}} \frac{\G{-\s}\G{\s+2\a}}{\G{\half+\a}}
  \2F1{-\s}{\s+2\a}{\half+\a}{\frac{1+Z}{2}} .
}
Here ${}_2F_1(a,b;c;z)$ is the Gauss hypergeometric function.
In general this function has a branch point at $Z=1$ and is cut
along the positive real axis $Z \in [1,+\infty)$. The time-ordered and
Wightman correlation functions of a massive scalar field $\phi_\s(X)$
are given by
\eq{ \label{eq:freeProp}
  \brak{0}{T \phi_\s(X_1)\phi_\s(X_2)}{0}_{\rm free}
  = \D_\s(X_1 \cdot X_2 - i \eps) ,
}
\eq{ \label{eq:freeWightman}
  \brak{0}{\phi_\s(X_1)\phi_\s(X_2)}{0}_{\rm free}
  = \D_\s(X_1 \cdot X_2  - i \eps s(X_1,X_2) ) ,
}
where in (\ref{eq:freeWightman}) the operator ordering is
enforced by $s(X_1,X_2) = +(-)$ if $X_1$ is in the future (past)
of $X_2$ (see, e.g., \cite{Goldstein:2003qf}).

At the level of free fields, one may in fact use any member of the
one-parameter family of 2-point functions found in
\cite{Allen:1985ux,Mottola:1984ar} to define a de Sitter-invariant
vacuum state. These other vacua are usually called Mottola-Allen (MA)
or $\a$ vacua. However, the non-trivial MA vacua do not satisfy the
Hadamard or Bunch-Davies criteria; in particular, their 2-point functions
i) have an additional singularity at antipodal points and ii) have an
additional negative frequency contribution to the singularity at coincident
points \cite{Bousso:2001mw}. As a result, only the Euclidean vacuum
extrapolates to the usual Minkowski vacuum in the flat space limit
\cite{deBoer:2004nd}. It has been difficult to find consistent extensions of
MA vacua to interacting theories -- see e.g.
\cite{Goldstein:2003qf,Goldstein:2003ut,Goldstein:2005re,
  Einhorn:2002nu,Einhorn:2003xb,Brunetti:2005pr}.
For these reasons we will discuss only the Euclidean vacuum in this work.

\section{Simple tree diagrams}
\label{sec:Euclidean}

We now proceed to analyze simple connected tree diagrams.  As noted in the introduction, we compute diagrams on Euclidean $S^D$ and analytically continue the results to de Sitter.  In particular, the Mellin-Barnes techniques used below provide representations of connected diagrams $\mathcal{V}_N(X_1,\dots,X_n,X_N)$ on $S^D$ in terms of the $N(N-1)/2$ embedding distances $Z_{ij} = X_i \cdot X_j$ relating the external points.  While the $Z_{ij}$ are not all independent for general $N,D$, it will often be convenient to use our Mellin-Barnes representation to extend the definition of $\mathcal{V}_N$ to a function of $N(N-1)/2$ independent variables $Z_{ij}$.  The analytic continuation can then be performed
by analytically continuing in each $Z_{ij}$ and evaluating
$Z_{ij} = X_i \cdot X_j$ for $N$ points $X_i$ in Lorentz-signature de Sitter space.

The only subtlety in the analytic continuations will be the presence of branch cuts.  As noted in section \ref{sec:free}, for the two-point function this amounts to choosing the appropriate $i\eps$ prescription to construct time-ordered or Wightman correlators, as desired.  Much the same is true of higher $N$-point correlators, though the specifics are more complicated to state.  However, since our only goal is to extract the asymptotics at large $Z_{ij}$, we need not be concerned with such details here.  The large $Z$ asymptotics are identical on both sides of each cut so that all analytic continuations satisfy the fall-off properties derived below. This means in particular that our results hold for both Wightman and time-ordered correlators.

\subsection{ The Green's function}
\label{sec:Delta}

It is convenient for our analysis to use a Mellin-Barnes integral
representation of the scalar Green's function on $S^D$. Mellin-Barnes representations
have proved to be quite useful in evaluating Feynman diagrams in
flat-space QFT (see, e.g., \cite{Smirnov:2004ym} for an introduction).
They are especially convenient for deriving asymptotic expansions
(see \S 4.8 of \cite{Smirnov:2004ym}), and it is for this
reason that we choose to use them here. We review some essential
information about Mellin-Barnes integrals in Appendix~\ref{app:MB};
further details can be found an any standard text on mathematical
methods.

Starting with the case $\sigma < \sigma_{cc} = \frac{1}{2} - \alpha$ and
$\alpha \ge \half$,  we may write the scalar Green's function
\eq{ \label{eq:DeltaMB}
  \D_\s(Z) = \frac{1}{(4\pi)^{\a+1/2}\GG{\half+\a+\s, \half-\a-\s}}
  \int_\nu \GG{-\s+\nu, \s + 2\a+\nu, -\nu, \half-\a-\nu}
  \left(\frac{1-Z}{2}\right)^\nu ,
}
where we use a condensed notation for products and ratios of $\Gamma$-functions:
\eq{
  \GGG{a_1,\,a_2,\,\dots,\,a_j}{b_1,\,b_2,\,\dots,\,b_k}
  := \frac{\G{a_1}\G{a_2}\cdots\G{a_j}}{\G{b_1}\G{b_2}\cdots\G{b_k}},
}
or merely $  \GG{a_1,\,a_2,\,\dots,\,a_j}$ for just a product.
In (\ref{eq:DeltaMB}) the symbol $\int_\nu \dots$ denotes a contour integral in
the complex $\nu$ plane. We take as implicit the measure $d\nu/2\pi i$.
The contour of integration is a straight line parallel to the imaginary
axis traversed from $-i\infty$ to $+i\infty$ anywhere within a region
called the ``fundamental strip'' (FS). In general we denote a fundamental
strip by its left and right boundaries $< l, r>$. For the Green's function
(\ref{eq:DeltaMB}) the fundamental strip is $< \s, \half-\a >$ which is non-empty due to the restriction $\s < \frac{1}{2} - \a$. The integrand
is analytic in $\nu$ within the FS; beyond the FS it has an infinite number
of poles due the Gamma functions. By convention we call poles
generated by Gamma functions $\Gamma(\cdots + \nu)$ left poles;
likewise, we call poles generated by Gamma functions $\Gamma(\cdots - \nu)$
right poles. The fundamental strip is the region between the left and
right poles. For this reason we do not generally need to write the FS
explicitly as it may be inferred from the Gamma functions of the integrand.

The asymptotic behavior of $\D_\s(Z)$ at large $|Z|\gg 1$ may be determined
by moving the contour to the left. The first of the two series of left
poles give the leading asymptotic terms:
\eqn{ \label{eq:DeltaLarge}
  \D_\s(|Z| > 1) &=& \frac{1}{4\pi^{\a+1}}
  \left\{
    \GG{-\s, \s+\a} (-2 Z)^\s + \GG{\s+2\a, -\s-\a} (-2 Z)^{-\s-2\a} \right\}
  \nn \\ & & \times
  \left[1 + O\left(Z^{-2}\right)\right] .
}
The asymptotic behavior for $|Z|$ near 1 is determined by moving the
contour to the right. When $D$ is odd,  $\a$ is an integer greater than or equal to $1$ and the leading
behavior is given by
\eqn{ \label{eq:DeltaSmall}
  \D_\s(|Z| < 1) &=& \frac{1}{(4\pi)^{\a+1/2}}
  \left\{
    \G{\a-\half} \left(\frac{1-Z}{2}\right)^{1/2-\a}
    + \GGG{\half-\a, -\s, \s+2\a}{\half-\a-\s, \half+\a+\s} \right\}
  \nn \\ & & \times
  \left[1 + O\left(1-Z \right)\right] .
}
When $D$ is even $\a = \half + n$, $n \in \Nat$ (where $\Nat$ are the non-negative integers) and the two sets of poles
overlap at $\nu \in \Nat$ yielding double-poles. As a result the pole at
$\nu=0$ gives a term with logarithmic behavior:
\eqn{ \label{eq:DeltaLog}
  \D_\s(|Z| < 1) &=&
    \frac{\G{n}}{(4\pi)^{n+1}} \left(\frac{1-Z}{2}\right)^{-n}
  \left[1 + O\left(1-Z \right)\right]
  \nn \\ & &
  - \frac{1}{(4\pi)^{n+1}} \GGG{1+\s+2n}{1+\s,1+n} \log(1-Z) + O(1)
}
(the first term is omitted when $n=0$).

When $\s > \s_{cc}$ the left-most right pole in (\ref{eq:DeltaMB}) lies to the left of the right-most left pole so that there are can be no straight contour
in between.  To arrive at an expression valid for all masses, consider again
the case $\sigma < \sigma_{cc}$ and move the contour in (\ref{eq:DeltaMB})
to the right past the first right pole at $\nu = \half -\a$ to obtain the
expression
\eqn{ \label{eq:DeltaShift}
  \D_\s(Z) &=& \frac{-1}{(4\pi)^{\a+1/2}}
  \int_\nu \GGG{-\s+\nu, \s + 2\a, -\nu, \frac{3}{2}-\a-\nu}{\half+\a+\s, \half-\a-\s}
  \frac{1}{(\nu-\half+\a)}
  \left(\frac{1-Z}{2}\right)^\nu \nn \\ & &
  + \frac{\G{\a-\half}}{(4\pi)^{\a+1/2}}\left(\frac{1-Z}{2}\right)^{1/2-\a} .
}
In the integral in the first line the contour lies in the interval
$(\max\{\s,\half-\a\}, \min\{0,\frac{3}{2}-\a\})$.  This interval is non-trivial for $\s < \frac{3}{2} - \a$ (since $\s < 0$), and (\ref{eq:DeltaShift}) is a valid representation of the propagator for any such $\s$. This process can be repeated
as needed so that one can then increase $\s$ as far into the
complementary series as desired. The asymptotic properties when $\s > \half -\a$ are
again given by (\ref{eq:DeltaLarge})-(\ref{eq:DeltaLog}). At conformal coupling
$\s = \half-\a$, only the residue term in (\ref{eq:DeltaShift}) survives:
\eq{
  \D_{\rm cc}(Z)
  = \frac{\G{\a-\half}}{(4\pi)^{\a+1/2}}\left(\frac{1-Z}{2}\right)^{1/2-\a} .
}

The behavior of the Green's function at large $M^2 \gg 1$ will be
important to our analysis. Starting with (\ref{eq:DeltaMB}) we
define
\eq{ \label{eq:psi}
  \psi_\s(\nu)
  := \frac{1}{(4\pi)^{\a+1/2}}
  \GGG{-\s+\nu,\s+2\a+\nu,\half-\a-\nu}{\half+\a+\s, \half-\a-\s},
}
so that the Green's function may be written
\eq{
  \D_\s(Z) = \int_\nu \psi_\s(\nu) \G{-\nu} \left(\frac{1-Z}{2}\right)^\nu .
}
At large $M^2 \gg 1$ the function $\psi_\s(\nu)$ has the asymptotic
behavior
\eq{ \label{eq:psiLargeM}
  \psi_\s(\nu) = \frac{M^{2\a-1+2\nu}}{(4\pi)^{\a+1/2}}\G{\half-\a-\nu}
  \left(1 + O\left(M^{-2}\right) \right) ,
}
and as a result the Green's function has the asymptotic behavior
\eq{ \label{eq:DeltaLargeM}
  \D_\s(Z) = \frac{M^{2\a-1}}{(4\pi)^{\a+1/2}}
  \int_\nu \GG{-\nu, \half-\a-\nu} M^{2\nu} \left(\frac{1-Z}{2}\right)^\nu
  \left(1+O\left( M^{-2} \right) \right) .
}
Note that (\ref{eq:DeltaLargeM}) contains no left poles; the left poles
of the original expression (\ref{eq:psi}) do not appear at any finite order
in the expansion in inverse powers of $M^2$. In the limit $M^2 \to \infty$
the inequality $|M^2(1-Z)/2| > 1$ holds
for any fixed $Z \neq 1$, and in this limit the contour in (\ref{eq:DeltaLargeM})
may be closed in the left half-plane giving $\D_\s(Z\neq 1) = O(M^{-4})$.
By examining the action of (\ref{eq:DeltaLargeM}) integrated against
a test function (represented as an MB integral) one may determine that (\ref{eq:DeltaLargeM}) is
equivalent to
\eq{
  \D_\s(Z) =
  \frac{1}{M^2} \frac{1}{{\rm vol}(S^{2\a})} \frac{\delta(Z-1)}{(1-Z^2)^{\a-1/2}}
  + O\left(M^{-4}\right) ;
}
the first few sub-leading terms are
\eqn{ \label{eq:DeltaLargeM2}
  \D_\s(Z) &=&
  \frac{1}{M^2} \frac{1}{{\rm vol}(S^{2\a})} \frac{\delta(Z-1)}{(1-Z^2)^{\a-1/2}}
  + \frac{1}{M^4} \frac{1}{{\rm vol}(S^{2\a})}
  \frac{\partial}{\partial Z} \left [\frac{\delta(Z-1)}{(1-Z^2)^{\a-1/2}} \right]
  \nn \\ & &
  + \frac{1}{M^6} \frac{1}{{\rm vol}(S^{2\a})}
  \frac{\partial^2}{\partial Z^2} \left [\frac{\delta(Z-1)}{(1-Z^2)^{\a-1/2}} \right]
  + O\left(M^{-8}\right) .
}
Of course, the expansion (\ref{eq:DeltaLargeM2}) follows from the fact
that the Green's function is the inverse of the Klein-Gordon operator using $\frac{\mathbbm 1}{\nabla^2 - M^2} = -  M^{-2}\frac{\mathbbm 1}{1 - \nabla^2/M^2} = - M^{-2}(1 + \nabla^2/M^2 + \ldots){\mathbbm 1}$.

\subsection{Pauli-Villars regularization}
\label{sec:PV}

Feynman diagrams containing loops in general contain UV divergences
which must be dealt with through the process of perturbative
renormalization. For our purposes it is convenient to use Pauli-Villars (PV)
renormalization \cite{bogoliubov:1980aa}. In PV regularization we
replace the original scalar Green's function $\D_\s(Z)$ with
the regularized function
\eq{
  \D_\s^{\rm reg}(Z) := \D_\s(Z) + \sum_{i=1}^{[D/2]} C_i \D_{\r_i}(Z) .
}
Here $[\dots]$ denotes the integer part. This function is nothing more than the
original Green's functions plus Green's functions of heavy particles with masses
$M_i^2 = -\r_i(\r_i+2\a)$. We take the masses $M_i^2$ to belong to the principal series so that  $\D_\s^{\rm reg}(Z)$ will decay for large $|Z| > 1$ at the same rate as $\D_\s(Z)$. The coefficients $C_i$ are bounded functions of the
$M_i^2$ chosen to make $\D_\s^{\rm reg}(Z)$ finite at $Z=1$; i.e., to cancel the UV-divergent terms in $\D_\s(Z)$
(including the logarithmic divergences that occur for even dimensions).
For example, for $D=2,3$ the PV-regularized Green's function is
\eq{
  \D_\s^{\rm reg}(Z) = \D_\s(Z) - \D_{\rho}(Z), \quad {\rm for\;} D=2,3
}
while for $D=4,5$ it is
\eq{
  \D_\s^{\rm reg}(Z) := \D_\s(Z) + C_1 \D_{\r_1}(Z) + C_2 \D_{\r_2}(Z) ,
  \quad {\rm for\;}D=4,5
}
where the coefficients satisfy
\eq{
  C_1+C_2 = -1, \quad C_1 M_1^2 + C_2 M_2^2 = - M^2 .
}
One may write similar expressions for any dimension (see e.g. \cite{bogoliubov:1980aa}) and, if desired, one may make further PV subtractions to ensure that $\D_\s^{\rm reg}(Z)$ is differentiable to any desired order at $Z=1$. Such additional subtractions are useful in dealing with either field-renormalization counter-terms or derivatively coupled theories.  Below, we assume for simplicity of notation that neither of these is present in our theory.  However, the analysis is identical in the presence of derivative couplings so long as one
assumes sufficient PV subtractions to have been made to render all diagrams finite at the desired order of perturbation theory\footnote{For theories that are power-counting renormalizable, one may fix the set of PV subtractions independent of the order in perturbation theory. On the other hand, non-renormalizable theories should be treated as effective theories. In this case, there is no harm in taking the regularization scheme (i.e., the set of PV subtractions) to depend on the order in perturbation theory to which one works.}.
In particular, detailed specification of these subtractions is not needed.

The cancellation of UV singularities has immediate implications for the Mellin-Barnes representation of the regulated propagators.  Since the short-distance expansion is determined by the location of the right-poles, and since right poles with ${\rm Re} \nu < 0$ give terms divergent at $Z=1$ (where the character of the divergence depends on the location of the pole),
all such right-poles must cancel; i.e., the fundamental strip for the regularized propagators may be extended to $<\s,0>$ without picking up any explicit pole terms of the sort that appeared in
(\ref{eq:DeltaShift}). It follows that for any $\s < 0$ we may write
the regularized Green's function as
\eq{ \label{eq:DeltaReg}
  \D_\s^{\rm reg}( Z ) = \int_{\nu}\psi^{\rm reg}_{\s}(\nu) \G{-\nu}
  \left(\frac{1-Z}{2}\right)^{\nu}
}
with
\eq{ \label{eq:psiReg}
  \psi^{\rm reg}_\s(\nu) := \psi_\s(\nu) + \sum_{i=1}^{[D/2]} C_i \psi_{\rho_i}(\nu) .
}
The function $\psi^{\rm reg}_\s(\nu)$ is analytic on the interval
$(\Re \s , \half)$ in odd dimensions and $(\Re \s , 1)$ in even
dimensions. Using the results in appendix \ref{app:MB} one may readily
show that the function $\psi_\s(\nu)$ -- and therefore $\psir_\s(\nu)$ as
well -- has the asymptotic behavior
\eq{
\label{eq:asPsi}
  |\psi_\s(x + i y)|
  = e^{-3 |y|/2}|y|^{-1+x}\left[1 + O\left(|y|^{-1}\right)\right]
  \quad {\rm for \;}|y| \gg 1 .
}
Furthermore, the integrand in (\ref{eq:DeltaReg}) has only a simple pole at
$\nu =0$ which insures that there is no logarithmic UV divergence.

The PV-regularized Green's function $\D_\s^{\rm reg}(Z)$ is a bounded
function of $Z$. Because the sphere is compact it follows that using
the regularized Green's function to compute correlation functions yields
regularized correlation functions that are bounded functions of
the embedding distances. The UV divergences of the original perturbation
series are recovered in the limit $M_i^2 \to + \infty$.
We consider theories which can be renormalized by subtracting local counter-terms with coefficients depending on the regulator masses $M_i$.  As remarked in footnote \ref{foot} above, one would expect this procedure to be equivalent (up to finite local counter-terms) to the renormalization prescription given in \cite{Hollands:2001fb}, and thus to define a fully covariant renormalized quantum field theory in the sense of \cite{Hollands:2008vx} whenever the flat-space limit is power-counting renormalizable.

\subsection{Single-vertex diagrams}
\label{sec:1V}

\begin{figure}
  \label{fig:1V}
  \includegraphics[]{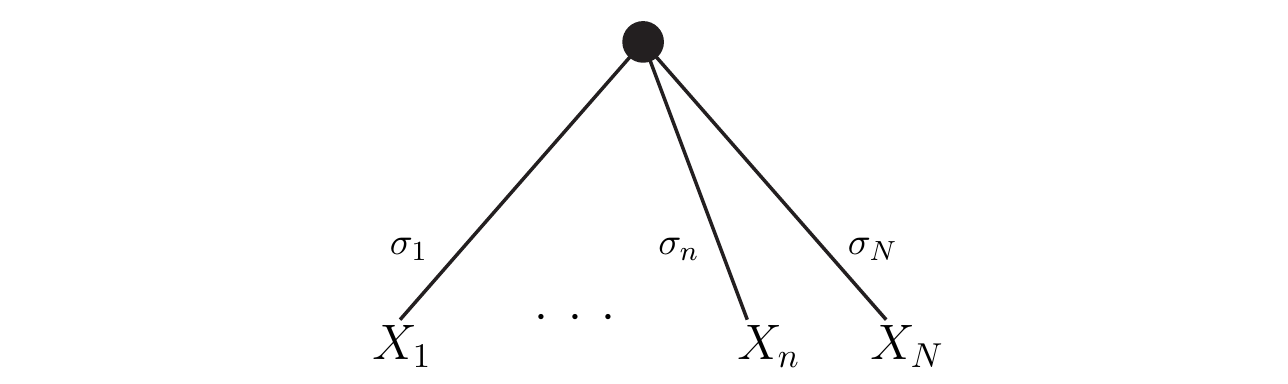}
  \caption{The single-vertex tree Feynman diagram.}
\end{figure}

In this section we compute the connected, single-vertex tree-level Feynman
diagram that arises lowest order in perturbation theory; see Fig.~3.
As stated in section \ref{sec:PV}, for simplicity of notation we assume below that there are derivative couplings.  However, the analysis in the presence of derivative couplings is essentially identical.

We find it convenient to first use the PV-regulated Green's functions $\D_\s^{\rm reg}(Z)$
for our computation and then to take the limit where the regulators are removed. While such regularization is not
in fact necessary for tree diagrams,  it has the convenient property that it
allows us to use the MB representation (\ref{eq:DeltaReg}) which treats all masses
uniformly.  Our discussion below involves a set of fields with mass parameters $\sigma_i$. Note that each $\sigma_i$ requires its own set of regulator masses $M_{ij}$, so removing the regulators is the limit $M_{ij} \rightarrow \infty$ (or $\rho_{ij} \rightarrow  \infty$).

The diagram in Fig.~3
is given by the expression
\eq{ \label{eq:1V}
  \cV_N(X_1,\dots,X_n,X_N)
  =
  \int_Y \Dr_{\s_1}(X_1\cdot Y) \cdots \Dr_{\s_n}(X_n \cdot Y) \Dr_{\s_N}(X_N \cdot Y) .
}
Here $Y$ is a unit vector and $\int_Y\dots$ denotes an integral over
$S^D$.  To compute the right-hand side we first expand the Green's
functions $\Dr_{\s_i}(X_i\cdot Y)$ according to (\ref{eq:DeltaReg}):
\eq{
  \Dr_{\s_i}(X_i\cdot Y) = \int_{\nu_i}\psir_{\s_i}(\nu_i) \G{-\nu_i}
  \left(\frac{1-X_i\cdot Y}{2}\right)^{\nu_i} .
}
After inserting $N$ copies of this into (\ref{eq:1V}) the integral over
$Y$ becomes
\eq{
  \cM_N := \int_Y \left(\frac{1-X_1\cdot Y}{2}\right)^{\nu_1}\cdots
  \left(\frac{1-X_n\cdot Y}{2}\right)^{\nu_n}
  \left(\frac{1-X_N\cdot Y}{2}\right)^{\nu_N} .
}
This master integral is performed in Appendix~\ref{app:MN}; the result
is
\eqn{ \label{eq:MN}
  \cM_N &=& \frac{(4\pi)^{\a+1/2}}
  {\GG{-\nu_1,\dots,-\nu_n,-\nu_N,1+2\a+\sum\nu_i}}
  \int_{(a)} \Bigg\{
  \left(\frac{1-X_1 \cdot X_2}{2}\right)^{a_{12}}
  \cdots
  \left(\frac{1-X_n \cdot X_N}{2}\right)^{a_{nN}}
  \nn \\ & &
  \GG{-a_{12},\dots,-a_{nN}, A_1 - \nu_1,\dots, A_N - \nu_N,
    \half +\a + \sum \nu_i - \sum a_{ij}} \Bigg\} .
}
Here $\int_{(a)}\dots$ denotes an integral over $N(N-1)/2$ integration
variables $a_{ij}$. The $a_{ij}$ are labelled according to the corresponding
embedding distance $X_i\cdot X_j$. We use the shorthand $A_i = \sum_{j=1}^N a_{ij}$.
The integration contours lie between their respective left and right poles.
After performing the shift of variables $\nu_i \to \nu_i + A_i$ we
obtain
\eqn{
  \label{eq:V1ans}
  & &\C{\phi_{\s_1}(X_1)\cdots\phi_{\s_n}(X_n)\phi_{\s_N}(X_N)}
  \nn \\
  &=& \int_{(a)} \left\{ \left(\frac{1-X_1\cdot X_2}{2}\right)^{a_{12}} \cdots
  \left(\frac{1-X_n\cdot X_N}{2}\right)^{a_{nN}}
  \GG{-a_{12},\dots,-a_{nN}}
  V_N(a) \right\}
}
with
\eqn{ \label{eq:VN}
  V_N(a) &=& (4\pi)^{\a+1/2} \int_{[\nu]}\Bigg\{
  \GG{-\nu_1,\dots,-\nu_N}
  \psir_{\s_1}(A_1+\nu_1)\cdots\psir_{\s_N}(A_N+\nu_N)
  \nn \\ & & \phantom{(4\pi)^{\a+1/2} \int_{[\nu]}\bigg\{\;}
  \GGG{\half+\a+\sum \nu_i + \sum a_{ij}}{1+2\a+\sum \nu_i + 2\sum a_{ij}}
  \Bigg\} .
}

Our main task is to determine the fundamental strip of each $a_{ij}$
variable. The Gamma functions in (\ref{eq:V1ans}) restrict the
fundamental strip of each $a_{ij}$ variable to satisfy $\Re a_{ij} < 0$.
To further determine the FS we must determine where the
function $V_N(a)$ ceases to be analytic in the $a_{ij}$.
When all $a_{ij}$ satisfy $\Re a_{ij} \le 0$
the function $V_N(a)$ imposes no further restriction on
the right side of the fundamental strips. Because of the symmetry
of the diagram we need only study one variable in detail, say $a_{12}$.
As a function of $a_{12}$ the function $V_N(a)$ has left poles at
\eq{
  a_{12} = \s_1 - A_1' - n, \quad
  a_{12} = \s_2 - A_2' - n, \quad
  a_{12} = -\half-\a-\sum{}' a_{ij} - n .
}
In this expression $n \in \Nat$, $A_1' = A_1 - a_{12}$, etc.,
and $\sum{}'a_{ij} = \sum a_{ij} - a_{12}$. We conclude
that the FS of $a_{12}$ is
\eq{
\label{eq:a12tree}
  a_{12}:\;< \max\left\{ \s_1- A_1', \s_2 - A_2' ,
    -\half-\a-\sum{}' a_{ij} \right\}, 0>.
}
Analogous statements hold for the remaining $a_{ij}$.   In (\ref{eq:a12tree}) and below we take the operation max to select the greatest real part of any of its arguments. Note in particular that since the regulator masses $M_{ij}$ lie in the principle series (so that ${\rm Re} \ \rho_{ij} = - \alpha$ is fixed) the allowed strip (\ref{eq:a12tree}) is independent of the values chosen for the regulator masses $M_{ij}$, though it does depend on the precise locations chosen for the other contours.

We can use our knowledge of the fundamental strips of the $a_{ij}$ variables
to bound the behavior of the diagram $\cV_N$ at large embedding distances
$Z_{ij}$. For example, consider the case $|Z_{12}|\gg 1$ and all other
$Z_{ij} \neq 1$. We are free to arrange the $a_{ij}$ integration contours such
that all $a_{ij}$ except $a_{12}$ are fixed satisfying $\Re a_{ij} = -\eps$
where $\eps$ is an infinitesimal positive constant. In this configuration
the FS of $a_{12}$ becomes
\eq{
  a_{12}:\;< \max\left\{\s_1,\s_2\right\} + O(\eps), 0 > .
}
 We can therefore move the $a_{12}$ integration contour to
$a_{12} = \max\left\{\s_1,\s_2\right\} + O(\eps)$. In this
configuration it becomes clear that the diagram decays at least as fast
as $|Z_{12}|^{\max\{\s_1,\s_2\}+O(\eps)}$. More generally we may say that
when any embedding distance satisfies $|Z_{ij}|\gg 1$ the diagram decays at least
as fast as $|Z_{ij}|^{\s_{\rm max}+O(\eps)}$ where
$\s_{\rm max} = \max\{\s_1,\dots,\s_N\}$ and infinitesimal $\eps > 0$.

The diagram $\cV_N$ provides the connected part of the
PV-regulated N-point correlation function
$\C{\phi_{\s_1}(X_1)\cdots\phi_{\s_n}(X_n)\phi_{\s_N}(X_N)}$
to lowest order in perturbation theory.   Our primary goal is to
determine the behavior of such connected correlators
when the operators are taken to large separations, so that several embedding distances $Z_{ij}$ become large. From the discussion
above it follows that the connected PV-regulated correlator decays at least as fast
as $|Z|^{\s_{\rm max}+O(\eps)}$, where $|Z|$ is the largest embedding distance
between operators. In practice the diagram may decay much more
rapidly.

In order to show that the unregulated diagrams have the same
IR behavior, we must take the limit $M_{i}^2 \to \infty$ where the regulator masses
become large. The key step is to recall, as noted below (\ref{eq:a12tree}), that  the allowed locations of the $a_{ij}$ contours are independent of the regulator masses $M_{ij}$.  We may therefore investigate the large $M_{ij}$ behavior by inserting the asymptotic expansion (\ref{eq:psiLargeM}) for the $\psi_{\rho_{ij}}(A_i+\nu_i)$, associated with the propagators for the PV regulator masses, into (\ref{eq:VN}) with the $a_{12}$ contour fixed at any location allowed by (\ref{eq:a12tree}) (and  analogously for the other $a_{ij}$).
To leading order, all dependence on the regulator masses is in factors of the form
$(\rho_1)^{2\a-1+2A_1+2\nu_1}$. The particular power law depends on the location of the $\nu_i$ contours, and the most favorable behavior is obtained by taking the $\nu_i$ contours to be as far to the left as possible.  With this in mind, taking into account certain relevant poles, it is straightforward to analyze the large $M_{ij}$ behavior.  The leading term is independent of $M_{ij}$ and is obtained by simply replacing every $\psi^{\rm reg}_\sigma$ with the unregulated $\psi_\sigma$; i.e., just by the unregulated expression.  Sub-leading terms are suppressed by powers of $M_{ij}^{-2}$ and can be neglected.  Since the unregulated $\psi_\s$ also satisfy (\ref{eq:asPsi}) at large imaginary $\nu_i$, the ${\cal O}(1)$ Mellin-Barnes integral can be analyzed in the usual way to find asymptotic behaviors at large $|Z_{ij}|$ dictated by the locations of the $a_{ij}$ contours; i.e., by (\ref{eq:a12tree}) and its analogues.  Thus the large $|Z_{ij}|$ behavior of the $M_{ij} \rightarrow \infty$ limit satisfies the same bounds we derived at finite $M_{ij}$.  In particular, the limiting diagram decays at least as fast as $|Z|^{\s_{\rm max}+O(\eps)}$, where $|Z|$ is the largest embedding distance between operators.

\section{General Diagrams}
\label{sec:loops}

In this section we analyze connected Feynman diagrams containing loops.
We again use the PV-regulated propagators of section \ref{sec:PV}.  For simplicity of notation we again assume that there are no derivative couplings or field-renormalization counter-terms.  However, the analysis with
derivative couplings or field-renormalization counter-terms is essentially identical so long as sufficient PV subtractions have been made as described in section \ref{sec:PV}.

At the technical level, the key step will be to show in section
\ref{sec:proof} that all diagrams have a Mellin-Barnes representation
of the following form:

\eqn{ \label{eq:form}
  & &\mathcal{V}_N(X_1,\dots,X_n,X_N)
  \nn \\
  &=& \int_{(a)} \left\{ \left(\frac{1-X_1\cdot X_2}{2}\right)^{a_{12}}
    \cdots
    \left(\frac{1-X_n\cdot X_N}{2}\right)^{a_{nN}}
    \GG{-a_{12},\dots,-a_{nN}}
    V_N(a) \right\}  ,
}
where the function $V_N(a)$ satisfies the following requirements:
\begin{enumerate}
\item
  $V_N(a)$ is analytic when all $a_{ij}$ are contained within
  the region given by the set of restrictions
  \eq{ \label{eq:region}
    \Re a_{ij} \in (\s_{\rm max} - \cP_{ij}(a'), 0] .
  }
  Here $\s_{\rm max}$ is the real part of the mass parameter of the lightest field
  participating in the diagram and $\cP_{ij}(a')$ is
  a polynomial function of all the $\Re a_{kl}$ variables except $\Re a_{ij}$
  (hence the prime) and has non-negative coefficients.
\item
  When the $a_{ij}$ are contained in the region (\ref{eq:region})
  the function $V_N(a)$ decays at large $|\Im a_{12}|\gg 1$ at least
  as rapidly as
  \eq{ \label{eq:req2}
    \left| V_N(x+iy,a_{13},\dots,a_{nN})\right| \propto e^{-\pi |y|/2} |y|^{-1+x}, \quad {\rm for\;} |y|\gg 1 ,
  }
  and likewise for the other $a_{ij}$.
\end{enumerate}

However, let us first discuss the implications of this form and show that it leads to exponentially decaying correlators as desired.

\subsection{Implications of our Mellin-Barnes representation}
\label{sec:implications}

To begin, note that the requirement
(\ref{eq:req2}) ensures that each integral in (\ref{eq:form}) converges so long as \emph{no} embedding distance is equal to unity, i.e. when
the diagram is evaluated away from coincident points. For any
$a_{ij} = x + i y$ the integrand in (\ref{eq:form}) is comparable at
large $|y| \gg 1$ to
\eq{
  e^{-\pi |y|+i\pi y} |y|^{3/2} \left|\frac{1-X_i\cdot X_j}{2}\right|^x,
}
and thus converges absolutely. To evaluate
$\cV(X_1,\dots,X_n,X_N)$ at coincident points we must move some of the
contours into the right half-plane. For example, suppose we wish to
evaluate $\cV_N(X_1,\dots,X_n,X_N)$ at $X_1 = X_2$. To do so we first
move the $a_{12}$ contour into the right half-plane. In doing so
pick up a residue from the pole at $a_{12}=0$. From (\ref{eq:region}) it
follows that $V_N(a_{12}=0,\dots)$ is regular and so this pole is a
simple pole. Upon setting $X_1 \cdot X_2 = 1$ the remaining contour
integral, with $a_{12}$ (slightly) in the right half-plane vanishes,
leaving just the residue:
\eqn{ \label{eq:VNat1}
  \mathcal{V}_N(X_2,X_2,\dots,X_n,X_N)
  &=& \int_{(a')} \bigg\{
  \left(\frac{1-X_2\cdot X_3}{2}\right)^{a_{13}+a_{23}} \cdots
  \left(\frac{1-X_2\cdot X_N}{2}\right)^{a_{1N}+a_{2N}}
  \nn \\ & & \phantom{\int_{(a')} \bigg\{ \; }
  \left(\frac{1-X_3\cdot X_4}{2}\right)^{a_{34}} \cdots
  \left(\frac{1-X_{n}\cdot X_N}{2}\right)^{a_{nN}}
  \GG{-a_{13},\dots,-a_{nN}}
  \nn \\ & & \phantom{\int_{(a')} \bigg\{ \;}
  V_N(0,a_{13},\dots,a_{nN}) \bigg\} .
}
Here $\int_{(a')}\dots$ denotes that there is no $a_{12}$ integral.

In fact, it turns out that the term on the right-hand side of (\ref{eq:VNat1})
may be written in form (\ref{eq:form}), i.e. $\cV_{N-1}(X_2,\dots,X_N)$.
Said differently, a function $\cV_{N+K}(X_1,\dots,X_N,X_{N+1},\dots,X_{N+K})$
when evaluated at $X_{N+1}=\dots=X_{N+K}=Y$ is itself a function of the form
$\cV_{N+1}(X_1,\dots,X_N,Y)$.
For example, let us consider when $K=2$. Following the
procedure outlined above equation (\ref{eq:VNat1}) we have
\eqn{ \label{eq:pinch}
  & & \cV_{N+2}(X_1,\dots,X_N,Y,Y) =
  \int_{(a')} \bigg\{
  \left(\frac{1-X_1\cdot X_2}{2}\right)^{a_{12}} \cdots
  \left(\frac{1-X_n\cdot X_N}{2}\right)^{a_{nN}}
  \nn \\ & & 
  \left(\frac{1-X_1\cdot Y}{2}\right)^{a_{1,N+1}+a_{1,N+2}} \cdots
  \left(\frac{1-X_N\cdot Y}{2}\right)^{a_{N,N+1}+a_{N,N+2}}
  \GG{-a_{12},\dots,-a_{N,N+2}}
  \nn \\ & & V_{N+2}(a_{12},\dots,a_{N,N+2},0)
  \bigg\} .
}
In this expression the prime in the $(a')$ below the integral means that there
is no $a_{N+1,N+2}$ integration.
The integrand in (\ref{eq:pinch}) is still analytic with respect to
the remaining $a_{ij}$ in the region
given by (\ref{eq:region}). It follows that after a few cosmetic
changes we may write (\ref{eq:pinch}) in the form of (\ref{eq:form}).
Let us re-label the variables
$a_{i,N+2} \to c_i$ (here $i=1,\dots,N$), then shift variables
$a_{i,N+1} \to a_{i,N+1}-c_i$; (\ref{eq:pinch}) becomes
\eqn{
  & &\mathcal{V}_{N+2}(X_1,\dots,X_N,Y,Y)
  \nn \\
  &=& \int_{(a)} \bigg\{
    \left(\frac{1-X_1\cdot X_2}{2}\right)^{a_{12}}
    \cdots
    \left(\frac{1-X_n\cdot X_N}{2}\right)^{a_{nN}}
    \left(\frac{1-X_1\cdot Y}{2}\right)^{a_{1,N+1}}
    \left(\frac{1-X_N\cdot Y}{2}\right)^{a_{N,N+1}}
    \nn \\ & & \phantom{\int_{(a)} \bigg\{ \;}
    \GG{-a_{12},\dots,-a_{N,N+1}}
    V^{\rm new}_{N+1}(a) \bigg\}  .
}
In this expression the integral is over the variables
$a_{12},\dots,a_{N,N+1}$ and $V^{\rm new}_{N+1}(a)$ is
given by
\eqn{
  V^{\rm new}_{N+1}(a) &:=& \frac{1}{\GG{-a_{1,N+1},\dots,-a_{N,N+1}}}
  \int_{[c]}
  \bigg\{
  \GG{c_1-a_{1,N+1},\dots,c_N-a_{N,N+1},-c_1,\dots,-c_N}
  \nn \\ & &
  V_{N+2}(a_{12},\dots,a_{nN},c_1-a_{1,N+1},\dots,c_N-a_{N,N+1},
  c_1,\dots,c_N,0)
  \bigg\} .
}
In this expression $\int_{[c]}\dots$ denotes contour integration
over $c_1,\dots,c_N$. These integrals are guaranteed to converge
so long as the $a_{ij}$ are within the region for which the integrand
of (\ref{eq:pinch}) is analytic.
Although this expression is rather complicated, it is easy
to verify that this function satisfies requirements (1) and (2) using
the asymptotics described in appendix \ref{app:MB}.  The same analysis
may be performed for any $K > 1$ with the same conclusion: the function
$\cV_{N+K}(X_1,\dots,X_N,Y,\dots,Y)$ is of the form of a function
$\cV_{N+1}(X_1,\dots,X_N,Y)$ given by (\ref{eq:form}).

The last and most important consequence of the form (\ref{eq:form})
is that the function $\mathcal{V}_N(X_1,\dots,X_n,X_N)$ decays exponentially when
evaluated at large embedding distances. For example, suppose
$|X_1\cdot X_2| \gg 1$. A bound on the decay of $\cV_N(X_1,\dots,X_n,X_N)$
can be found in the same manner as in the previous section. Let all integration
contours except that of $a_{12}$ be located at $\Re a_{ij} = -\eps$. From
(\ref{eq:region}) it follows that in this configuration $a_{12}$
has a fundamental strip at least as large as
\eq{
  a_{12}:\;< \s_{\rm max} + O(\eps), 0 > ,
}
so $\cV_N(X_1,\dots,X_n,X_N)$ decays at least as fast as
$(X_1 \cdot X_2)^{\s_{\rm max}+\epsilon}$ for any $\epsilon > 0$.

Furthermore, suppose that removing some vertex results in a disconnected diagram, and suppose also that one of the resulting connected components contains none of the original external legs.  Then this piece contributes only an overall multiplicative constant (which is finite at finite regulators masses $M_{ij}$) to the diagram and does not affect the large $Z$ behavior.  One may therefore remove such pieces from the diagram when computing $\sigma_{\rm max}$ above.  We refer to this process as ``trimming," so that the trimmed version of a given diagram has all such pieces removed.

Obviously, the same result also holds for the other embedding distances.
From this result it follows that the connected part of a PV-regulated $N$-point
function -- which may be described to any order in perturbation
theory by diagrams of the form $\cV_N$ -- decays when any two operators
are taken to be separated by a large distance $Z$ at least
as fast as $|Z|^{\s_{\rm max}+O(\eps)}$, where $\s_{\rm max}$ is the (real part of the) largest $\s$ that appears in any trimmed diagram that contributes to the correlator.

\subsection{Proof of the desired Mellin-Barnes representation}
\label{sec:proof}

The proof that all diagrams can be written in the form
(\ref{eq:form}) is through induction. One constructs a diagram
vertex by vertex, beginning with a single-vertex tree diagram.
We have already seen that single-vertex diagrams have MB integral representations of the
required form. Thus one simply needs to show that, upon adding a vertex
to an existing diagram with the form (\ref{eq:form}), the new diagram is again of the form (\ref{eq:form}). We show this below.

The process of adding a new vertex to an existing diagram
is shown schematically in Fig~4.
Starting with
an $(N+K)$-legged diagram
$\cV_{N+K}(X_1,\dots,X_N,X_{N+1},\dots,X_{N+K})$, one attaches a new vertex to the $K \ge 1$ external legs $X_{N+1},\dots,X_{N+K}$.
One then  attaches to the new vertex $(M-N)$ new external legs so that
the new diagram is an $M$-legged diagram:
\eqn{ \label{eq:VM}
  \cV_M(X_1,\dots,X_M) &=&
  \int_Y  \cV_{N+K}(X_1,\dots,X_N,Y,\dots,Y)
  \Dr_{\s_{N+1}}(X_{N+1}\cdot Y)\cdots
  \Dr_{\s_{M}}(X_{M}\cdot Y) . \nn \\
}

This procedure generates all diagrams in which no propagator has both of its ends on the same vertex.  But adding such one-link loops simply multiplies any diagram by factors of $\Dr_{\s}(Y \cdot Y)$, which are just finite constants due to our PV regularization, and which are readily absorbed into the definition of $\cV_M$.  It thus remains only to show that the diagrams generated by the above process satisfy requirements (1) and (2)  associated with (\ref{eq:form}).

\begin{figure}
  \label{fig:add}
  \includegraphics{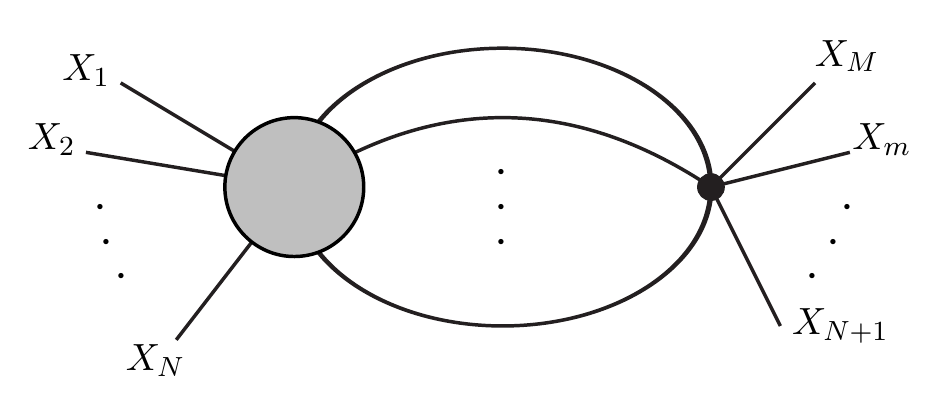}
  \caption{The process of adding a new vertex to an existing diagram.}
\end{figure}

Note that $K \ge 1$ in order for the diagram to be connected. Following the
discussion in section \ref{sec:implications}, since $K \ge 1$ we know that
$\cV_{N+K}(X_1,\dots,X_N,Y,\dots,Y)$
can be written in the form of some $\cV_{N+1}(X_1,\dots,X_N,Y)$. Inserting this
(\ref{eq:VM}) becomes
\eqn{ \label{eq:VM2}
  \cV_M(X_1,\dots,X_M) &=&
  \int_Y  \cV_{N+1}(X_1,\dots,X_N,Y)
  \Dr_{\s_{N+1}}(X_{N+1}\cdot Y)\cdots
  \Dr_{\s_{M}}(X_{M}\cdot Y) . \nn \\
}

It is convenient to define $n= N-1$ and $m=M-1$.
The integral (\ref{eq:VM}) can be computed in essentially the
same manner as the single-vertex diagram section \ref{sec:Euclidean}.
We begin by expressing both $\cV_{N+1}(X_1,\dots,X_N,Y)$ and the
regulated Green's functions in terms of their MB integral representations:
\eqn{
  & &\mathcal{V}_{N+1}(X_1,\dots,X_N,Y)
  \nn \\
  &=& \int_{(a)} \int_{\nu_1}\cdots \int_{\nu_N}
  \bigg\{
    \left(\frac{1-X_1\cdot X_2}{2}\right)^{a_{12}}
    \cdots
    \left(\frac{1-X_n\cdot X_N}{2}\right)^{a_{nN}}
    \left(\frac{1-X_1\cdot Y}{2}\right)^{\nu_1}
    \cdots
    \left(\frac{1-X_N\cdot Y}{2}\right)^{\nu_N}
    \nn \\ & &
    \GG{-a_{12},\dots,-a_{nN},-\nu_1,\dots,-\nu_N}
    V_{N+1}(a_{12},\dots,a_{nN},\nu_1,\dots,\nu_N)
  \bigg\}  ,
}
\eq{
  \Dr_{\s_i}(X_i\cdot Y) = \int_{\nu_i}\psir_{\s_i}(\nu_i) \G{-\nu_i}
  \left(\frac{1-X_i\cdot Y}{2}\right)^{\nu_i} .
}
Here it is important to keep track of notation. In the first equation we have
relabelled $a_{i,N+1} \to \nu_i$, $i=1,\dots,N$, so that the remaining
$N(N-1)/2$ $a_{ij}$ variables run $a_{12},\dots,a_{nN}$. In the second
expression $i$ runs $i=N+1,\dots,M$. Inserting these into (\ref{eq:VM2})
we then integrate over $Y$ using the master integral $\cM_{M}$
(see (\ref{eq:MN})). After performing a shift of integration variables
$\nu_i \to \nu_i + B_i$ (where $B_i = \sum_{j=1}^N b_{ij}$) we arrive at
\eqn{ \label{eq:VM3}
  & & \mathcal{V}_M(X_1,\dots,X_M)
  = \int_{(a)} \int_{(b)} \bigg\{
    \left(\frac{1-X_1\cdot X_2}{2}\right)^{a_{12}+b_{12}}
    \cdots
    \left(\frac{1-X_n\cdot X_N}{2}\right)^{a_{nN}+b_{nN}}
    \nn \\ & &
    \left(\frac{1-X_1\cdot X_{N+1}}{2}\right)^{b_{1,N+1}}
    \cdots
    \left(\frac{1-X_m\cdot X_{M}}{2}\right)^{b_{mM}}
    \GG{-a_{12},\dots,-a_{nN},-b_{12},\dots,-b_{mM}}
    V_M(a,b) \bigg\}  . \nn \\
}
In this expression the $b_{ij}$ run over all distinct pairs $ij$ (i.e., $b_{12},\dots,b_{mM}$) and
\eqn{
  V_M(a,b) &=& \int_{[\nu]} \Bigg\{
  \GGG{-\nu_1,\dots,-\nu_M,\half + \a + \sum \nu_i + \sum b_{ij}}
  {1 + 2\a + \sum \nu_i + 2 \sum b_{ij}}
  \nn \\ & &
  V_{N+1}(a_{12},\dots,a_{nN},B_1+\nu_1,\dots,B_N+\nu_N)
  \psir_{\s_{N+1}}(B_{N+1}+\nu_{N+1})\cdots
  \psir_{\s_{M}}(B_{M}+\nu_{M}) \Bigg\} . \nn \\
}

It is now straightforward to determine the region for which the
integrand in (\ref{eq:VM3}) is analytic in the integration variables.
The simplest variables to analyse are the $b_{ij}$ variables with
$N <i,j \le M$. For these variables the analysis is identical to
that performed for the single-vertex graph; the result is that
the integrand is analytic in the region
\eq{ \label{eq:bmM}
  {\rm Re} \ b_{mM} \in \;\left( \max\left\{\s_m - B_m', \s_M - B_M', - \half - \a - \sum{}' b_{ij}\right\},
  0 \right] .
}
As usual here the prime denotes that $b_{mM}$ is omitted from the
sums. For variables $b_{ij}$ with $1 \ge i \le N$ and $N < j \le M$
one finds
\eq{
  {\rm Re} \  b_{1M} \in \;\left( \max\left\{\s_{\rm max} - \cP_{1,N+1}(a,b), \s_M - B_M',
    - \half - \a - \sum{}' b_{ij}\right\},
  0 \right] .
}
Finally, let us determine the region for which the integrand is analytic
with respect to $a_{ij}$ while holding the $b_{ij}$ contours with $1 \le i,j \le N$
fixed to satisfy $\Re b_{ij}= -\eps$. In this configuration it is easy
to determine that the integrand is analytic when
\eq{ \label{eq:a12}
  {\rm Re} \ a_{12} \in \;\left( \s_{\rm max}-\cP_{12}(a,b)+O(\eps), 0\right] .
}
Therefore, we can perform the shift of variables $a_{ij} \to a_{ij}-b_{ij}$ in
order to get (\ref{eq:VM3}) in the form (\ref{eq:form}). We know that the
$b_{ij}$ integrals with $1 \le i,j \le N$ will converge in the region given
by (\ref{eq:bmM})-(\ref{eq:a12}). We see that (\ref{eq:bmM})-(\ref{eq:a12})
satisfy (\ref{eq:region}), and that all integrals converge sufficiently
rapidly to satisfy (\ref{eq:req2}). Thus we have shown that $\cV_{M}(X_1,\dots,X_M)$
is of the form \ref{eq:form}.

\subsection{The regulator limit $M_{ij}^2 \to \infty$}
\label{sec:LargeM}

Our analysis above is complete at the level of effective theories.  In that context, one keeps the regulators masses $M_{ij}$ finite and is careful to ask questions only about physics at energy scales much less than $M_{ij}$.  But for renormalizable theories one would like to do more and to remove the regulators by sending $M_{ij} \rightarrow \infty$ before taking the limit of large $|Z_{ij}|$.

Such questions are straightforward to address using our Mellin-Barnes representations. Note that, as with the tree diagrams discussed in section \ref{sec:1V}, we may study the large $M_{ij}$ limit holding fixed the locations of all contours, subject only to the conditions (\ref{eq:region}) found above. Suppose for the moment that we choose the couplings to be independent of the regulators masses $M_{ij}$.  Then all of the regulator-dependence lies in the functions
$\psi_{\rho}(\nu)$ associated with the regulator Green's functions and the coefficients $C_i$.  Note that each term in the asymptotic expansion (\ref{eq:psiLargeM}) of such functions at large $M_{ij}$  again decays exponentially away from the real axis (now roughly as $e^{-\pi |y|}$) fast enough for the arguments of sections \ref{sec:implications},\ref{sec:proof} to hold\footnote{In fact, such arguments require decay only as $e^{-\pi|y|/2}$ times an appropriate power law or faster.}. As a result, inserting the expansion (\ref{eq:psiLargeM}) into one of our Mellin-Barnes integrals (and also expanding the $C_i$) produces an asymptotic series in the masses $M_{ij}$, each of whose coefficients is again a Mellin-Barnes integral with the same contours and convergence properties as the original expression.

Of course, the above expansion will in general include positive powers of $M_{ij}$ as well as negative powers; these are just the expected ultra-violet divergences of the theory.  But let us suppose that by taking the coupling constants to depend on $M_{ij}$ in an appropriate way the $M_{ij} \rightarrow \infty$ limits of correlators become well-defined and finite, at least to some fixed order in perturbation theory.  This is precisely the assumption that the divergences can be cancelled by some set of $M_{ij}$-dependent counter-terms.  Since coupling constants are just overall multiplicative factors in each diagram, it is straightforward to take this extra dependence on the $M_{ij}$ into account. Expanding each coupling in an asymptotic series generates a new series, where each term is again a Mellin-Barnes integral of our standard form (and with the same placement of the contours).
This is true in particular of the term that is independent of the $M_{ij}$.  But this term gives the full $M_{ij} \rightarrow \infty$ limit, since
all terms involving positive powers of $M_{ij}$ must have cancelled in order to obtain a finite result.  The usual argument then implies that this term decays as $|Z|^{\sigma_{\rm max} + {\cal O} (\epsilon)}$ at large $|Z|$, where $\s_{\rm max}$ is the (real part of the) largest $\s$ that appears in any trimmed diagram that contributes to the correlator at this order.

\section{Discussion}
\label{sec:disc}

In the above work, we used Mellin-Barnes techniques to determine the asymptotics of Pauli-Villars regulated diagrams for massive scalar quantum field theories in de Sitter space.  We found that connected correlators fall off at large $|Z|$ at least as fast as does the Green's function for the lightest field in the (trimmed) diagram (up to corrections that grow less strongly than powers laws; e.g., factors of $\log|Z|$).  Due to the simple way in which changing the PV regulator masses interacted with the Mellin-Barnes expressions, it was straightforward to show that the same results hold in the $M_{ij} \rightarrow \infty$ limit in which the regulators are removed, independent of the details of any counter-terms required.  A similar analysis using Mellin-Barnes techniques should also be possible in the context of dimensional regularization.

As described in the introduction, it follows that the interacting Hartle-Hawking vacuum is an attractor state in the sense of \cite{Marolf:2010zp} for local correlators at any order of perturbation theory.   Our results hold for all masses $M^2 > 0$ for which a free Euclidean vacuum exists and for arbitrary interactions, with non-renormalizable theories being treated as effective theories.  While for simplicity of notation the calculations were presented only for non-derivative couplings, no significant changes are required to analyze derivatively-coupled theories and (as usual) derivatives can only strengthen the fall-off at large $Z$.    It would be very interesting if our results could be extended to the massless case $M^2 =0$ following e.g. the approach of \cite{Rajaraman:2010xd}, which introduced a new form of perturbation theory on $S^D$.

Some readers may be concerned by our use of Euclidean techniques.  But on general grounds the Hartle-Hawking state should be a valid quantum state.  In particular,
the analytically continued correlators satisfy the Lorentz-signature Schwinger-Dyson equations.  Furthermore, the de Sitter analogue \cite{Schlingemann:1999mk} of the Osterwalder-Schr\"ader construction implies that the Hartle-Hawking state lives in a positive-definite Hilbert space whenever the Euclidean correlators satisfy reflection-positivity.  This in turns holds at least formally whenever the Euclidean action is bounded below, and has been rigorously shown in $D=2$ dimensions for standard kinetic terms and polynomial potentials; see e.g. \cite{Glimm:1987ng}. In such cases, it remains only to ask how the Hartle-Hawking  state relates to other states of interest; e.g, perhaps the state defined by the standard in-in perturbation theory in the expanding cosmological patch of $dS_D$.  This question will be investigated in detail in \cite{Higuchi:2010aa}, where it will be shown that these two states agree for massive scalar fields.

\begin{acknowledgements}
The authors thank David Berenstein, Cliff Burgess, Steven B. Giddings,
Atsushi Higuchi, Stefan Hollands, Alexander Polyakov, Mark Srednicki,
and Richard Woodard for enlightening discussions.
This work was supported in part by the US National Science Foundation
under grants PHY05-55669 and PHY08-55415 and by funds from the University
of California.
\end{acknowledgements}

\appendix

\section{Mellin-Barnes integrals}
\label{app:MB}

We write a generic Mellin-Barnes integral as \footnote{
  This discussion follows closely the discussion in
  \cite{Bateman:1955}.
}
\eq{ \label{eq:genericMB}
  f(Z) = \int_\nu
  \GGG{a_1+A_1\nu,\dots, a_m + A_m \nu, b_1-B_1,\dots, b_n-B_n \nu}
  {c_1+C_1 \nu,\dots c_p+C_p\nu, d_1-D_1 \nu,\dots, d_q-D_q \nu}
  (Z)^\nu ,
}
where the measure $d\nu/2\pi i$ is implicit and the contour is
a straight line parallel to the imaginary axis, traversed from
$-i\infty$ to $+i\infty$, lying between the left and right poles.
The convergence of the integral (\ref{eq:genericMB}) is governed by
the behavior of the integrand at large $|\Im\,\nu|$. This behavior
can be determined from the well-known asymptotic behavior of the
Gamma function:
\eq{
  \lim_{|y|\to\infty} \G{x+i y}
  = (2\pi)^{1/2} e^{-\frac{\pi}{2}|y|} |y|^{x-1/2}
  \left[1 + O(y^{-1}) \right] .
}
Let us assume that the all $A_i$, $B_i$, $C_i$, $D_i$ are positive
and define
\eqn{
  E &=& \sum_{i=1}^m A_i + \sum_{i=1}^n B_i
  - \sum_{i=1}^p C_i - \sum_{i=1}^q D_i, \\
  F &=& \sum_{i=1}^m A_i - \sum_{i=1}^n B_i
  - \sum_{i=1}^p C_i + \sum_{i=1}^q D_i, \\
  G &=& \Re\left[\sum_{i=1}^m a_i + \sum_{i=1}^n b_i
  - \sum_{i=1}^p c_i - \sum_{i=1}^q d_i \right] + \half (-m-n+p+q), \\
  H &=& \prod_{i=1}^m (A_i)^{A_i} \prod_{i=1}^n (B_i)^{-B_i}
  \prod_{i=1}^p (C_i)^{-C_i} \prod_{i=1}^q (D_i)^{D_i} ,
}
and furthermore let $Z = R e^{i \Phi}$ and $\nu = x + i y$.
With this notation the absolute value of the integrand behaves like
\eq{
  \exp \left[ -\Phi y - \frac{E \pi}{2} |y|\right]
  |y|^{F x + G} (R H)^{x}
}
as $|y|\to \infty$. From this we conclude that the integral
(\ref{eq:genericMB}) is absolutely convergent when
\begin{enumerate}
  \item $|\Phi| < E\pi/2$. The integral \ref{eq:genericMB}
    defines an analytic function of $Z$ for
    $|{\rm arg}\,Z| < {\rm min}\left(\pi,\frac{E\pi}{2}\right)$.
  \item $|\Phi| = E \pi /2$ and $Fx+G < -1$. The integral
    defines an analytic function for all $Z$.
\end{enumerate}
See \cite{Bateman:1955} for further details.

\section{Calculation of $\cM_N$}
\label{app:MN}

In this appendix we compute the integral
\eq{\label{eq:I}
  \cM_N := \cM(\nu_1,\dots,\nu_N)
  = \int_Y \left(\frac{1-X_1 \cdot Y}{2}\right)^{\nu_1}
  \cdots \left(\frac{1-X_n \cdot Y}{2}\right)^{\nu_n}
  \left(\frac{1-X_N \cdot Y}{2}\right)^{\nu_N}
}
with $n=N-1$. Rather than directly evaluating (\ref{eq:I}) we
instead consider the integral
\eq{ \label{eq:A}
  \calA(\a_1,\dots,\a_n) := \int_Y \left[
    \a_1 \left(\frac{1-X_1 \cdot Y}{2}\right) + \cdots +
    \a_n\left(\frac{1-X_n \cdot Y}{2}\right)
    + \left(\frac{1-X_N \cdot Y}{2}\right)\right]^\l ,
}
where $\a_i$ are arbitrary real parameters and $\l$ is a complex
number with $\Re\l < 0$.
The quantities $\calA$ and $\cM$ may be related in a simple way.
To do so we use a standard Mellin-Barnes formula:
\eqn{ \label{eq:MBexpansion}
& & (A_1 + \cdots + A_n + A_N)^\l \nn \\ & &
= \frac{1}{\G{-\l}}
\int_{u_1} \cdots \int_{u_n} \GG{-\l+\sum_{i=1}^n u_n,\,-u_1,
  \dots,\,-u_n} (A_1)^{u_1}\cdots (A_n)^{u_n} (A_N)^{\l-\sum  u_i} .
\nn \\
}
Inserting (\ref{eq:MBexpansion}) in (\ref{eq:A}) yields
\eqn{ \label{eq:A_MB}
  \calA(\a_1,\dots,\a_n) &=& \frac{1}{\G{-\l}}
  \int_{u_1} (\a_1)^{u_1} \cdots \int_{u_n} (\a_n)^{u_n} \bigg\{
  \GG{-\l+\sum_{i=1}^n u_i,\,-u_1,\,\dots,\,-u_n}
  \nn \\ & & \phantom{\frac{1}{\G{-\l}}
    \int_{u_1} (\a_1)^{u_1} \cdots \int_{u_n} (\a_n)^{u_n}\bigg\{ }
 \cM\left(u_1,\dots,u_n,\l-\sum_{i=1}^n u_i\right) \bigg\}.
}
Written this way $\cM$ is one factor of the Mellin transform
of $\calA$.

Let us now return to (\ref{eq:A}) and integrate over $Y$. We use
the formula
\eq{ \label{eq:beta}
  \eta^\l = \frac{i^{-\l}}{\G{-\l}} \int_0^\infty d\beta \,\beta^{-1-\l}
  e^{-i\beta \eta}
}
to write $\calA$ as
\eq{ \label{eq:A_int}
  \calA(\a_1,\dots,\a_n) = \frac{(2i)^{-\l}}{\G{-\l}}
  \int_0^\infty d\beta\,\beta^{-1-\l}
  \exp\left[-i\beta \left(1 + \sum_{i=1}^n \a_i\right)\right]
  \int_Y e^{+i\beta V Y}
}
where $V = \a_1 X_1 + \cdots + a_n X_n + X_N$. The integral
over $Y$ can be written in terms of the Bessel function:
\eq{ \label{eq:Yint}
  \int_Y e^{-i \beta V Y}
  = (2\pi)^{\a+1} \frac{J_{\a}(\beta |V|)}{(\beta |V|)^{\a}} .
}
The Bessel function may be written as a Mellin-Barnes integral
\eq{ \label{eq:BesselJ_MB}
  J_\nu(z) = \int_\mu \GGG{-\mu}{1+\nu+\mu}\left(\frac{z}{2}\right)^{\nu+2\mu};
}
inserting (\ref{eq:BesselJ_MB}) into (\ref{eq:Yint}) yields
\eq{ \label{eq:Yint2}
  \int_Y e^{-i \beta V Y}
  = 2\pi^{\a+1} \int_\mu \GGG{-\mu}{1+\a+\mu}
  \left(\frac{\beta^2 V^2}{4}\right)^\mu .
}
After inserting (\ref{eq:Yint2}) into (\ref{eq:A_int}) we may
integrate over $\beta$ using the inverse of (\ref{eq:beta})
\eq{
  \int_0^\infty d\beta \,\beta^{-1-\l} e^{-i\beta \eta}
  = \frac{\G{-\l}}{i^{-\l}} (\eta - i0)^\l .
}
Convergence of the integral over $\beta$ requires $\Re(\l-2\mu) < 0$.
The result is
\eqn{ \label{eq:A_mu}
  \calA(\a_1,\dots,\a_n) = \frac{2^{1-\l}\pi^{\a+1}}{\G{-\l}} \int_\mu
  \GGG{-\mu,\,2\mu-\l}{1+\a+\mu} (2i)^{-2\mu}(V^2)^\mu
  \left(1+\sum_{i=1}^n \a_i\right)^{\l-2\mu} .
}

Next we perform a number of manipulations in order to tidy
up (\ref{eq:A_mu}). First note that
\eqn{
  V^2 &=& \a_1^2 + \cdots + \a_n^2 + 1 +
    \a_1 \a_2 X_1 X_2 + \cdots + \a_n X_n X_N \nn \\
  &=& \left(1+\sum_{i=1}^n \a_i\right)^2
    + 2 \a_1 \a_2(X_1 X_2 - 1) + \cdots
    + 2 \a_n (X_n X_N - 1) .
}
It is convenient to use \ref{eq:MBexpansion} to write
\eqn{
  (V^2)^\mu &=& \frac{1}{\G{-\mu}} \int_w \bigg\{\GG{-\mu+w,\,-w}
  \left(1+\sum_{i=1}^n \a_i\right)^{2(\mu-w)}
  \nn \\ & & \phantom{\frac{1}{\G{-\mu}} \int_w\bigg\{}
  \left[2\a_1\a_2(X_1 \cdot X_2-1) + \cdots + 2\a_n(X_nX_N-1)\right]^w
  \bigg\} .
}
Inserting this into (\ref{eq:A_mu}) yields
\eqn{
  \calA(\a_1,\dots,\a_n) &=& \frac{2^{1-\l}\pi^{\a+1}}{\G{-\l}}
  \int_\mu \int_w \bigg\{ \GGG{2\mu-\l,\,-\mu+w,\,-w}{1+\a+\mu}
  (2i)^{-2\mu}\left(1+\sum_{i=1}^n \a_i\right)^{\l-2w} \nn \\
  & & \phantom{\frac{2\pi^{\a+1}}{\G{-\l}} \int_\mu \int_w \bigg\{}
  \left[ 2\a_1 \a_2 (X_1 X_2-1)+\cdots + 2 \a_n (X_n X_N-1) \right]^w
  \bigg\}
}
We can now integrate over $\mu$. First we use the Gamma function
duplication formula
\eq{
  \GG{x,\,x+\half} = 2^{1-2x}\sqrt{\pi} \G{2x}
}
on the Gamma function $\G{2\mu-\l}$; second we use the Gauss summation
formula \cite{Bateman:1955} written here as a Mellin-Barnes integral:
\eq{
  \int_\mu \GGG{a+\mu,\,b+\mu,\,d-\mu}{c+\mu}e^{\pm i \pi \mu}
  = e^{\pm i \pi d} \GGG{a+d,\,b+d,\,c-a-b-d}{c-a,\,c-b} ,
}
valid for $\Re(c-a-b-d)>0$. Cleaning up we have
\eqn{ \label{eq:Aalmost}
  \calA(\a_1,\dots,\a_n) &=&
  \frac{2^{1+2\a}\pi^{\a+1/2}}{\GG{-\l,\,1+2\a+\l}}
  \int_w \bigg\{ \GG{2w-\l,\,\half+\a+\l-w,\,-w}
  \nn \\ & &
  \left(1+\sum_{i=1}^n \a_i\right)^{\l-2w}
  \left[\a_1\a_2 \left(\frac{1-X_1 \cdot X_2}{2}\right) + \cdots
    + \a_n \left(\frac{1-X_nX_N}{2}\right) \right]^w \bigg\} .
  \nn \\
}

The next series of steps is simple but rather cumbersome
to transcribe. We expand both the term in parentheses
and the term in square brackets in (\ref{eq:Aalmost}) using
the Mellin-Barnes expansion (\ref{eq:MBexpansion}). Within
the parentheses there are $n+1$ terms, so the Mellin-Barnes expansion
of this quantity has $n$ integrations. Likewise, the term in square brackets
has $N(N-1)/2$ terms so the Mellin-Barnes expansion of this quantity
has $N(N-3)/2$ integrations. After performing some shifts in the
integration variables (taking care not to shift a contour through a pole)
and relabelling we obtain the following expression:
\eqn{ \label{eq:Afinal}
  & &
  \calA(\a_1,\dots,\a_n) =
  \nn \\ &=&
  \frac{2^{1+2\a}\pi^{\a+1/2}}{\GG{-\l,\,1+2\a+\l}}
  \int_{\mu_1} (\a_1)^{\mu_1} \cdots \int_{\mu_n} (\a_n)^{\mu_n}
  \Bigg\{
  \nn \\ & &
  \int_{h_{12}} \cdots \int_{h_{nN}} \Bigg\{
  \left(\frac{1-X_1 \cdot X_2}{2}\right)^{h_{12}} \cdots
  \left(\frac{1-X_n\cdot X_N}{2}\right)^{h_{nN}}
  \GG{-h_{12},\dots,-h_{nN}}
  \nn \\ & &
  \GG{\sum h_{1i}-\mu_1,\,\dots,\,\sum h_{ni}-\mu_n,
    \,\sum h_{Ni}-\l+\sum_{i=1}^n \mu_i,\, \half+\a+\l-\sum h_{ij}}
  \Bigg\} \Bigg\}.
  \nn \\
}
In this expression there is a total of $n$ integration variables
$\mu_1,\dots,\mu_n$ and $N(N-1)/2$ variables $h_{ij}$. The latter
are labelled such that each factor of $(1-X_i \cdot X_j)/2$ is
raised to the power $h_{ij}$.

The convergence of each Mellin-Barnes integral may be evaluated
using the technique described in Appendix~\ref{app:MB}. Each integral
converges absolutely for all $(1-X_i \cdot X_j)/2 \neq 1$. The expression
(\ref{eq:Afinal}) defines a single-valued function of the inner
products $X_i \cdot X_j$ for all complex values of $X_i \cdot X_j$
away from the cuts $X_i \cdot X_j \in [1,\infty)$.

Both (\ref{eq:A_MB}) and (\ref{eq:Afinal}) equate $\calA$ with an
$n$-fold Mellin transform with parameters $\a_i$. It is easy to see
that the integration contours of the two expressions -- those of the $u_i$ in
the former expression and $\mu_i$ in the latter expression -- may be
taken to be traversed in the same places in their respective complex
planes. Now recall that the Mellin inversion theorem states that for
a given choice of integration contour the Mellin transform of a function
is unique \cite{Bateman:1955}. It follows that we may identify the integrands and equate
$u_1 = \mu_1,\; \dots,\; u_n = \mu_n$.
The final step is to relabel
\eq{
  \mu_i = u_i \to \nu_i, \quad {\rm for\;} i=1,\dots,n , \quad
  \l \to \nu_N + \sum_{i=1}^n {\nu_i}, \quad h_{ij} \to a_{ij},
}
which yields
\eqn{\label{eq:I_MB}
  & & \cM(\nu_1,\dots,\nu_N)
  \nn \\ & & = \frac{(4\pi)^{\a+1/2}}
  {\GG{-\nu_1,\dots,-\nu_N,1+2\a+\sum\nu_i}}
  \int_{(a)} \Bigg\{
  \left(\frac{1-X_1 \cdot X_2}{2}\right)^{a_{12}}
  \cdots
  \left(\frac{1-X_n \cdot X_N}{2}\right)^{a_{nN}}
  \nn \\ & & \phantom{==}
  \GG{-a_{12},\dots,-a_{nN}, A_1 - \nu_1,\dots, A_N - \nu_N,
    \half +\a + \sum \nu_i - \sum a_{ij}} \Bigg\} .
}
In this expression $A_i := \sum_{j=1}^N a_{ij}$.


\bibliography{./bibliography}

\end{document}